\documentclass[conference]{IEEEtran}
\IEEEoverridecommandlockouts

\usepackage{cite}
\usepackage{amsmath,amssymb,amsfonts}
\usepackage{algorithmic}
\usepackage{graphicx}
\usepackage{textcomp}
\usepackage{xcolor}
\usepackage{hyperref}
\usepackage{comment}
\usepackage{enumitem}
\usepackage{ulem}
\usepackage{booktabs}
\usepackage{tabularx}
\usepackage{footnote}
 \usepackage{soul} 
\usepackage{url}
\usepackage{hyphenat}
\usepackage{balance}
\usepackage{multirow}
\usepackage[listings]{tcolorbox}
\usepackage[skip=0pt, font=small]{caption}

\def\BibTeX{{\rm B\kern-.05em{\sc i\kern-.025em b}\kern-.08em
    T\kern-.1667em\lower.7ex\hbox{E}\kern-.125emX}}
\begin{document}

\title{Are LLMs Ready for Anti-Pattern Detection in Microservice Architectures?}

\author{\IEEEauthorblockN{Anonymous Authors}}

\author{
\vspace{-1em}
\hspace*{-1.35em}
\IEEEauthorblockN{
\begin{tabular}{ccc}
\begin{tabular}[t]{c}
 Marco De Luca\\
\textit{DIETI}\\
\textit{University of Naples Federico II}\\
Naples, Italy\\
marco.deluca2@unina.it
\end{tabular}
&
\begin{tabular}[t]{c}
 Domenico Amalfitano\\
\textit{DIETI}\\
\textit{University of Naples Federico II}\\
Naples, Italy\\
domenico.amalfitano@unina.it
\end{tabular}
&
\begin{tabular}[t]{c}
Porfirio Tramontana\\
\textit{DIETI}\\
\textit{University of Naples Federico II}\\
Naples, Italy\\
ptramont@unina.it
\end{tabular}
\\[6.3em]
\multicolumn{3}{c}{%
\begin{tabular}[t]{cc}
\begin{tabular}[t]{c}
Anna Rita Fasolino\\
\textit{DIETI}\\
\textit{University of Naples Federico II}\\
Naples, Italy\\
fasolino@unina.it
\end{tabular}
\end{tabular}%
}
\\
\end{tabular}
}
\vspace{-1em}
}

\maketitle

\begin{abstract}
Microservice systems are prone to recurrent architectural anti-patterns (APs) that hinder maintainability, evolvability, and operational quality. Most existing AP detection approaches rely on static analysis and handcrafted rules, which can be effective but are often tool-dependent, limited to explicitly encoded detection logic, and difficult to adapt to heterogeneous repositories. In this paper, we investigate whether large language models (LLMs) are ready to support architectural anti-pattern detection in microservice architectures through a prompt-based analysis pipeline over static repository artifacts. We evaluate three general-purpose LLMs on a curated benchmark of microservice repositories annotated with 16 architectural anti-patterns, and compare their performance against the state-of-the-art static-analysis tool MARS using a uniform evaluation protocol based on precision and recall. Our results show that LLMs can provide useful support for anti-pattern detection, achieving competitive performance on several anti-patterns, especially when the relevant evidence is local, heterogeneous, or semantically rich. At the same time, they exhibit clear limitations on anti-patterns that require explicit structural or cross-service dependency evidence, where static analysis remains more reliable. These findings suggest that LLMs are not yet a replacement for traditional analyzers, but already represent a promising complementary aid for architectural assessment in microservice systems.

\end{abstract}


\section{Introduction}
Microservice architectures (MSAs) have become increasingly popular due to their scalability, modularity, and support for independent service evolution. However, as these systems grow in size and complexity, architectural anti-patterns can emerge, leading to maintainability issues, reduced evolvability, and operational inefficiencies. Detecting such anti-patterns early is critical for preserving system quality and preventing architectural decay. Traditionally, microservice anti-pattern detection relied on static analysis techniques that inspect source code, configuration files, and structural dependencies to identify design issues. Tools such as MARS focus on architectural anti-patterns by analyzing service-level relations and deployment-related artifacts \cite{Tighilt2023}, while other approaches target more localized microservice smells within individual services \cite{MSANose2020}. Although effective, these approaches are typically rule-based and rely on explicitly encoded detection logic, which may reduce their flexibility when dealing with heterogeneous repositories or anti-patterns whose symptoms are spread across multiple artifacts.

Recent advances in Large Language Models (LLMs) open a new perspective for software architecture analysis. By processing source code and configuration artifacts in natural language form, LLMs can reason over heterogeneous evidence and potentially identify anti-pattern instances even when they are not captured by rigid handcrafted rules. This makes them promising candidates for supporting anti-pattern detection in MSAs, especially in cases where the relevant signals are implicit, dispersed, or semantically rich.

Despite this potential, there is still limited empirical evidence on the actual effectiveness of LLMs for architectural anti-pattern detection in microservice systems, and on how their behavior compares with established static-analysis tools. In particular, it remains unclear for which types of anti-patterns LLMs can provide reliable support, where they fall short, and whether they should be seen as alternatives to or complements of traditional analyzers.

This work investigates the use of LLMs for architectural anti-pattern detection in MSAs from static repository artifacts. We evaluate three general-purpose LLMs on a benchmark of microservice systems annotated with 16 architectural anti-patterns, and compare their precision and recall against the state-of-the-art static-analysis tool MARS. Through this comparison, we aim to understand the strengths and limitations of LLM-based detection and to provide evidence on how LLMs can complement traditional static-analysis techniques in the detection of microservice architectural anti-patterns.

The remainder of the paper is organized as follows. Section \ref{sec:rel} presents the related work on architectural anti-patterns in microservice systems. Section \ref{sec:methodology} describes the methodology of our study. Section \ref{sec:results} reports the experimental results. Section \ref{sec:ap-category-takeaways} discusses the main findings and their implications. Section \ref{sec:threats} presents the threats to validity. Finally, Section \ref{sec:conclusion} concludes the paper and outlines future work.

\section{Related Work}
\label{sec:rel}

This section presents the main lines of research relevant to our study. We first outline how architectural anti-patterns in microservice architectures have been defined and classified in the literature. We then discuss the main techniques and tools proposed for MSA analysis and anti-pattern detection. Finally, we consider AI-based approaches for identifying problematic behaviors and anti-pattern-related issues in microservice systems, highlighting the gap addressed by our work.

\subsection{MSA Architectural Anti-patterns}

The adoption of microservice architectures has introduced new design challenges, leading to the identification of specific code smells and architectural anti-patterns. These issues often stem from violations of core microservice principles such as loose coupling, bounded contexts, and domain-driven design.

Code smells are generally defined as symptoms in source code or design that may indicate deeper problems \cite{fowler1999refactoring}. They do not necessarily correspond to defects, but rather to design issues that can hinder maintainability, scalability, or evolvability if left unresolved. In microservices, typical examples include God Service, where a service grows too large and accumulates multiple responsibilities, and improper communication patterns, such as excessive synchronous calls, chatty APIs, or inconsistent messaging protocols. In contrast, anti-patterns are recurring poor solutions to common design problems \cite{brown1998antipatterns}, capturing flawed practices that negatively affect system quality. In MSAs, they often arise at a higher architectural level, for example Shared Persistence, where multiple services access the same database schema, Cyclic Service Dependencies, where services depend on each other in a loop, and Hardcoded Configuration or Secrets, which reduce flexibility and security.

However, the distinction between code smells and anti-patterns is not always sharp. Several studies classify both in MSAs \cite{Taibi2019MicroservicesAntiPatterns}, \cite{Tighilt2020}, but their taxonomies often overlap and the terminology is sometimes used inconsistently \cite{Meissner2021EAsLiT}. For instance, God Service may be treated either as a service-level code smell or as an architectural anti-pattern depending on the adopted viewpoint. This ambiguity makes careful interpretation of existing taxonomies essential when analyzing microservice quality \cite{8354414}. A comprehensive classification of microservice architectural anti-patterns is provided in \cite{Cerny2023MicroserviceAntiPatterns}. Based on a tertiary study, it consolidates prior literature into a catalog of 58 disjoint anti-patterns organized into five categories: Intra-Service Design, Inter-service Decomposition, Service Interaction, Security, and Team Organization. This classification offers a useful reference for both practitioners and researchers concerned with microservice quality.

\subsection{Techniques and Tools for MSA analysis and AP Detection}

Several approaches have been proposed for the architectural analysis of MSAs. These techniques can be broadly classified into categories that focus on different perspectives and data sources \cite{Bushong21}. The most established approach is \textit{static analysis}, which inspects source code without execution to identify structural flaws through metrics such as complexity, coupling, and cohesion \cite{ALFAYEZ2023107147, BALDASSARRE2020106377}. Schneider et al. \cite{Cerny_tool} provide a comprehensive empirical comparison of static analysis tools for architecture recovery in microservice applications. \textit{Dynamic analysis}, instead, observes the system at runtime and helps detect issues such as synchronization problems, bottlenecks, or cascading failures by monitoring performance, latency, and interactions \cite{Offline_Mining, maruf2022telemetry}. \textit{Model-based analysis} relies on abstract architectural models, such as dependency matrices or stochastic models, to reason about properties including reliability, availability, and maintainability \cite{Mendonca20, McZara20}. \textit{Graph-based analysis} represents architecture as a graph to study dependencies, communication flows, and service relationships, using either static artifacts or runtime traces \cite{BRANDON20, liu19}. Finally, the \textit{analysis of development team behavior} has also been used to reveal technical problems in MSAs by exploiting indicators such as code ownership, change frequency, and author dispersion \cite{CodeOwnershipPrinciples2024, CodeOwnershipAISecurity2023, AlignmentOwnershipContribution2019, SourcegraphVision2023}. Several studies apply these approaches specifically to MSA anti-pattern detection. Static analysis is adopted by MARS, which detects 16 microservice anti-patterns \cite{Tighilt2023}, and by MSANose, which detects up to eleven microservice code smells \cite{MSANose2020}. Unlike MARS, which operates at the system level by analyzing inter-service dependencies and architectural constraints, MSANose focuses on the internal design of individual microservices. Matar et al. \cite{Matar2023} exploit static analysis of design models to identify anti-patterns affecting microservice performance. Pattern-based analysis, built on static inspection, has also been proposed to detect recurring design structures or rule violations associated with known anti-patterns \cite{Pigazzini2020, Marquez19, Walker20}. However, dynamic analysis remains a major trend in microservices because of their polyglot nature and decentralization. Approaches based on telemetry and service dependency graphs, combining logging, tracing, and metrics, have been used to flag potential anti-patterns \cite{maruf2022telemetry}. H\"ubener et al. \cite{Hubener2022} propose automatic anti-pattern detection from execution traces, while Parker et al. \cite{Parker2023} discuss runtime detection and visualization mechanisms in a mapping study.

\subsection{AI-based Detection of MSA Anti-Patterns}

AI techniques are increasingly used to detect problematic behaviors in microservice systems that often correspond to architectural anti-patterns, such as bottlenecks, chatty services, and cascading failures. Although much of this work is framed as anomaly detection, it often targets the same structural and runtime issues architects classify as anti-patterns.

A first line of work applies deep learning to traces and graphs. Approaches based on deep Bayesian networks \cite{Ping2020}, Graph Neural Networks combined with LSTMs or PU-learning, and diffusion convolutional recurrent networks \cite{10490038} model invocation traces as sequences or dynamic graphs. These methods learn normal behavior and detect anomalous call chains with high precision and F1, outperforming earlier trace-based techniques \cite{Chen2023}.

A second group uses autoencoders and hybrid deep models, such as AE+LSTM, convolutional autoencoders, and TCN+VAE with causal inference \cite{Shahini2024, xing2025multi}, to learn latent trace representations. Deviations in the latent space are then used to signal anomalies or faults, often improving accuracy, reducing false positives, and shortening troubleshooting time.

Other studies rely on classical machine learning and federated or multi-task learning, using supervised MLPs and ensemble models over metrics and traces to classify anomalies at service or trace level \cite{Kha2023}.

Finally, some works explicitly target architectural anti-patterns by constructing dependency graphs from telemetry data \cite{10803303} or combining distributed tracing with rule-based analysis to identify issues such as bottlenecks or excessive coupling \cite{Hubener2022}. Still, most existing approaches focus on generic anomaly detection rather than explicitly modeling architectural anti-patterns, which highlights a gap in dedicated AI-based techniques for MSA anti-pattern detection.

\section{Research Methodology}
\label{sec:methodology}

The goal of our study is to explore the potential of Large Language Models (LLMs) to analyze code, configuration files, and the structural organization of microservice architectures, assessing their capability to detect anti-patterns in Microservice architectures (MSAs). We are not interested in exploring the LLM capabilities to infer anti-patterns based on information obtained by dynamic analysis of the architecture.
In particular, we aim not only to evaluate LLMs’ detection accuracy but also to compare their effectiveness with established static-analysis tools, such as \textsc{MARS} \cite{Tighilt2023}, which is known for identifying a wide range of architectural anti-patterns. This formulation allows us to investigate whether LLMs can complement or even enhance traditional static-analysis approaches in uncovering anti-patterns in MSAs.

The study investigates the following Research Questions:

\begin{enumerate}[label=\hspace{.2cm}RQ\arabic*:, leftmargin=.44in]
    \item \textit{How effective are LLMs at detecting anti-patterns in MSAs?}
    \item \textit{How do LLMs compare against state-of-the-art static analysis tools for anti-pattern detection in MSAs?}
\end{enumerate}

In the following, we describe the selected dataset of anti-patters, considered LLMs, evaluation metrics, and the experimental procedure adopted to assess our approach.

\subsection{Data Set of Anti-Patterns}

Unlike code smells, for which several publicly available repositories exist \cite{ds1, ds2, ds3}, to the best of our knowledge, there are no public repositories specifically collecting MSA anti-patterns. However, we can refer to research works proposing tools for detecting anti-patterns in MSAs and reporting microservice-based applications together with the corresponding datasets of anti-patterns. 

An overview of recently proposed tools, together with the set of anti-patterns they are able to detect, is reported in \cite{MSANose2020}.
As this paper shows, most reported tools are capable of detecting only a limited subset of anti-patterns. The most relevant tools in terms of coverage are \textsc{MSANose} \cite{MSANose2020} and MARS \cite{Tighilt2023}. Nevertheless, MARS has been shown to detect not only all the anti-patterns covered by MSANose, with higher precision and recall, but also additional architectural anti-patterns beyond its scope.

In our study, we therefore focus on the 16 APs detectable by \textsc{MARS}, reported in Table~\ref{tab:ap_list} together with their definitions and classification into four categories according to the taxonomy proposed by Cerny et al.~\cite{Cerny2023MicroserviceAntiPatterns}. This choice was driven by the need to evaluate LLM-based detection under a common protocol with an existing AP detection tool, using the same repositories, anti-pattern definitions, ground truth, and precision/recall metrics. 

As ground truth, we rely on the curated AP labels reported in the original \textsc{MARS} study. These labels were not generated by \textsc{MARS} in our experiment; rather, they were adopted as an external reference benchmark to evaluate both the LLMs and the re-executed \textsc{MARS} baseline on the same repository subset. Starting from this dataset, we sampled the repositories by accounting for the input-size constraints of the LLMs considered in our study (i.e., the maximum number of tokens that an LLM can handle in its context window, which for GPT‑5.2 reaches up to 400k tokens\footnote{ChatGPT Doc: \url{https:openai.com/index/introducing-gpt-5-2/}}), ensuring that all models could be evaluated under comparable conditions. The final list of the thirteen selected repositories, together with their characterization, is shown in Table~\ref{tab:repo_characterization}.



\begin{table*}[t]
\footnotesize
\renewcommand{\arraystretch}{1.1}
\caption{Microservice Anti-pattern Definitions}
\label{tab:ap_list}
\centering
\resizebox{\textwidth}{!}{%
\begin{tabular}{p{0.22\textwidth} p{0.78\textwidth}}
\toprule
\multicolumn{1}{c}{\textbf{Anti-pattern}} & \multicolumn{1}{c}{\textbf{Definition}} \\
\midrule
\midrule
\multicolumn{2}{c}{\textbf{Inter-service Decomposition}} \\
\midrule

Wrong Cuts \textbf{(WC)} &
WC consists of microservices organised around technical layers (business, presentation, and data) instead of functional capabilities, which causes strong coupling among microservices and impedes the delivery of new business functions. \\ \hline

Cyclic Dependencies \textbf{(CD)} &
CD occurs when multiple microservices are circularly co-dependent and thus no longer independent, which goes against the very definition of microservices. \\ \hline

Shared Libraries \textbf{(SL)} &
SL relates to the sharing of libraries and files (e.g., binaries) by multiple microservices, which breaks their independence as they rely on a single source to fulfil their business function. \\ \hline

Shared Persistence \textbf{(SP)} &
SP happens when multiple microservices share a single database: they no longer own their data and cannot use the most suitable database technology for their business function. \\ \hline

\midrule
\multicolumn{2}{c}{\textbf{Intra-Service Design}} \\
\midrule
Mega-service \textbf{(MS)} &
MS is a microservice that provides multiple business functions. A microservice should be manageable by a single team and should pertain to a single business function. \\ \hline

Nano-service \textbf{(NS)} &
NS results from a too fine-grained decomposition of a system, i.e., when one business function requires many microservices to work together. \\ \hline

No API Versioning \textbf{(NAV)} &
NAV happens when no information is available about a microservice version, which can break changes and force backward compatibility when deploying updates. \\ 
\hline
\midrule

\multicolumn{2}{c}{\textbf{Service Interaction}} \\
\midrule

Hard-Coded Endpoints \textbf{(HE)} &
HE relates to URLs, IP addresses, ports, and other endpoints being hardcoded in the source code of microservices and/or configuration files, which interferes with load balancing and deployment. \\ \hline

No API Gateway \textbf{(NAG)} &
NAG occurs when consumer applications (mobile applications, etc.) communicate directly with microservices and must know how the whole system is decomposed, managing endpoints and URLs for each microservice. \\ \hline

Time Outs \textbf{(TO)} &
TO happens when timeout values are set and hard-coded in HTTP requests, which leads to unnecessary disconnections or delays. \\ \hline

No Health Check \textbf{(NHC)} &
NHC describes microservices that are not periodically health-checked. Unavailable microservices may not be noticed, leading to timeouts and other errors. \\ \hline
\midrule

\multicolumn{2}{c}{\textbf{Team-Organization}} \\
\midrule
Manual Configuration \textbf{(MC)} &
MC refers to configurations that must be manually pushed in some microservices. Since microservice-based systems evolve rapidly, their management should be automated, including their configuration. \\ \hline

No Continuous Integration \/ Continuous Delivery (\textbf{NCI}) &
Continuous integration and delivery are important for microservices to automate repetitive steps during testing and deployment. Not using CI/CD undermines the microservice architectural style, which encourages automation wherever possible. \\ \hline

Multiple Service Instances Per Host \textbf{(MSIPH)} &
MSIPH happens when multiple microservices are deployed on a single host (e.g., container, physical machine, virtual machine), which prevents their independent scaling and may cause technological conflicts inside the host. \\ \hline

Local Logging \textbf{(LL)} &
LL results from microservices having their own logging mechanism, which prevents aggregation and analysis of their logs and the monitoring and recovery of systems. \\ \hline

Insufficient Monitoring \textbf{(IM)} &
IM describes neglecting to record data on performance levels and failures of microservice-based systems that would be useful for maintenance purposes. \\
\bottomrule
\end{tabular}%
}
\end{table*}

\begin{table}[t]
\footnotesize
\renewcommand{\arraystretch}{0.95}
\caption{Selected repositories and size characterization.}
\centering
\resizebox{0.9\columnwidth}{!}{%
\begin{tabular}{l c c c}
\toprule
\textbf{Repo} & \textbf{\#Microservices} & \textbf{\#Files} & \textbf{\#LOC} \\
\midrule
Apollo~\cite{repo:apollo} & 9 & 68 & 29510 \\
TeaStore~\cite{repo:teastore} & 3 & 62 & 5073 \\
Spring Cloud  Movie~\cite{repo:springcloudmovie} & 4 & 33 & 885 \\
Freddy's BBQ~\cite{repo:freddysbbq} & 6 & 35 & 1752 \\
Piggymetrics~\cite{repo:piggymetrics} & 4 & 88 & 3176 \\
FTGO~\cite{repo:ftgo} & 9 & 257 & 8239 \\
SB Microservices~\cite{repo:oktadev-springboot-ms} & 2 & 4 & 116 \\
LakeSide Mutual~\cite{repo:lakesidemutual} & 9 & 424 & 89477 \\
Warehouse~\cite{repo:warehouse-microservice} & 6 & 222 & 4623 \\
Qbike~\cite{repo:qbike} & 5 & 77 & 2057 \\
Microservice Demo~\cite{repo:ewolff-microservice-demo} & 3 & 38 & 1766 \\
CQRS ~\cite{repo:cqrs-sampler} & 3 & 26 & 1028 \\
Micro Company~\cite{repo:micro-company} & 17 & 244 & 90315 \\
\bottomrule
\end{tabular}%
}
\label{tab:repo_characterization}
\end{table}

\subsection{LLM Selection}
We selected three state-of-the-art, general-purpose LLMs: \texttt{ChatGPT~5.2}, \texttt{Gemini~3 Pro}, and \texttt{Qwen~3.5 Plus}. 
These models were chosen to provide a representative snapshot of current high-performing systems from different model families, to reduce the risk that our findings depend on the specific behavior of a single provider architecture. 
Moreover, all three models are designed to handle long, structured inputs and to perform multi-step reasoning over heterogeneous software artifacts (e.g., source code and configuration files), which aligns with our goal of assessing LLM-based anti-pattern detection from static information only. 
By evaluating multiple LLMs under the same experimental protocol, we can observe how their detection behavior varies and compare it more fairly with established static-analysis tools such as \textsc{MARS}.

\subsection{Metrics}

To evaluate the effectiveness of LLMs for AP detection, we use two metrics: \textit{Precision} and \textit{Recall}~\cite{GoutteGaussier2005}.  To answer \textit{RQ1}, we compare each LLM's outputs directly against the Ground Truth reported in~\cite{Tighilt2023}, and compute precision and recall accordingly. 

To answer \textit{RQ2}, we independently re-executed the publicly available \textsc{MARS} tool on the same repository subset, using the released/base configuration and without AP- or repository-specific tuning. Both the LLM predictions and the outputs obtained from our \textsc{MARS} execution were then evaluated against the same GT using precision and recall.

\subsection{Experimental Procedure}
\begin{figure}[h!]
  \centering
  \includegraphics[width=\linewidth]{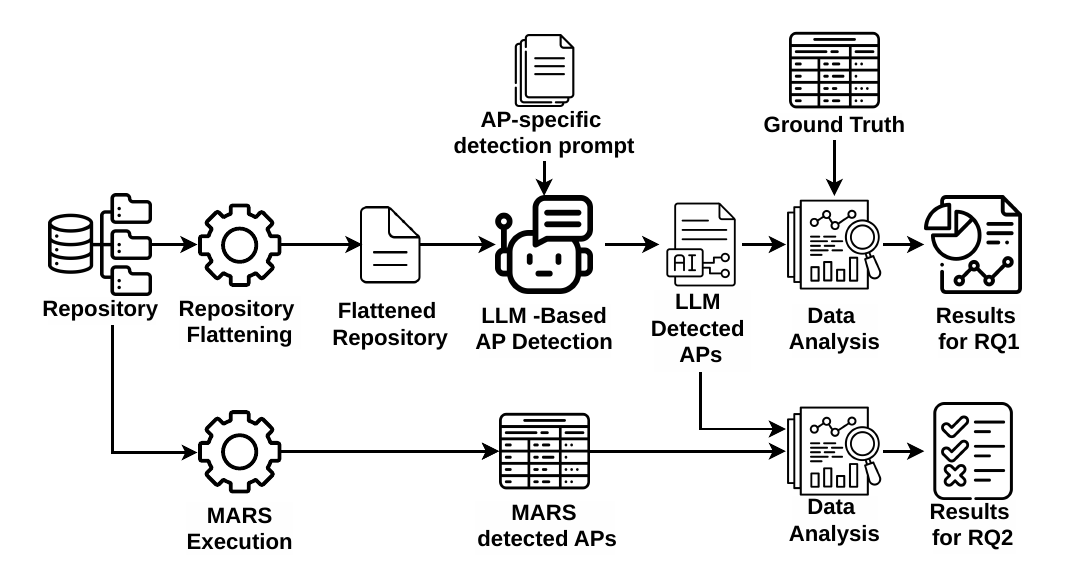}
  \caption{Experimental Procedure Pipeline}
  \label{fig:pipeline}
\end{figure}

Our experimental procedure consists of four main steps: (i) \textit{Repository Flattening}, where we convert each repository into a compact, LLM-consumable representation; (ii) \textit{Detection Prompt Execution}, where we run LLM-based anti-pattern (AP) detection through a parametric prompt; (iii) \textit{MARS Execution}, where we independently execute MARS on the same repository subset and set of APs; and (iv) \textit{Data Analysis}, where we evaluate both LLM and MARS predictions against the ground truth and compare their detection performance. Figure~\ref{fig:pipeline} provides an overview of the experimental procedure pipeline.

\paragraph{Repository Flattening}
To provide the LLM with a consistent input representation and to reduce parsing overhead, we first \textit{flatten} each repository using Repomix\footnote{Repomix: \url{https://github.com/yamadashy/repomix}.}, a tool that packs an entire codebase into a single AI-friendly textual artifact, including a synthetic view of the project structure and the concatenation of repository files. 
Repomix supports inclusion/exclusion rules; we configured these filters to discard non-relevant artifacts (e.g., images, datasets, documentation assets, and generated build outputs) and retain only source code and configuration files  (e.g., \texttt{Dockerfile}, \texttt{docker-compose.yml}, \texttt{pom.xml}, and \texttt{application.yml}).
Moreover, to reduce noise and token usage, we enabled Repomix comment removal so that the flattened representation focuses on implementation and configuration content rather than inline documentation.

\paragraph{Detection Prompt Execution}
In the second step, we provide the LLM with the flattened repository along with an AP-specific \textit{detection prompt}\footnote{Prompt executed on February 2026.}. 
For each anti-pattern, we run an independent inference by pairing the same flattened repository with the corresponding AP prompt. Moreover, for size-related APs (\textit{Nano-service} and \textit{Mega-service}), the input is augmented with a per-service size summary (LOC per microservice).

The output of this step is tailored to the anti-pattern under analysis. For APs that are defined as system-level properties, the prompt asks the model to return a binary decision (present/absent) for the whole repository. For APs that can be attributed to specific microservices, the prompt instead requires the model to report the list of affected microservices (possibly empty), providing an explicit microservice-level attribution of the detected AP instances.

\paragraph{MARS Execution}
To support the comparison with a static-analysis baseline, we executed MARS tool on the same repository subset and for the same set of MARS-detectable anti-patterns considered in our study. We independently re-executed the publicly available tool using its released configuration, without AP- or repository-specific tuning. This choice reflects the way an external researcher could run the tool from the public release and avoids giving MARS an additional configuration advantage with respect to the LLM-based pipeline.

\paragraph{Data Analysis}
Finally, we perform two complementary analyses. First, we compare the outputs produced by each LLM with the ground truth labels adopted from the original MARS study. To ensure a consistent evaluation, we distinguish between APs assessed at the repository level and APs attributed to specific microservices, aligning the LLM predictions with the corresponding reference labels. Second, we compare the outputs produced by the LLMs with those obtained from our independent execution of MARS tool on the same repositories.
The resulting matches and mismatches form the basis for the quantitative analysis presented in the results section.

\subsection{Prompt Design}
Our AP-detection prompt follows a parametric design, combining (i) a shared task-level template reused across anti-patterns and (ii) an AP-specific section defining the target anti-pattern and the observable evidence required in source code and configuration artifacts. This structure ensures consistency while allowing controlled specialization. The design was guided by the prompt engineering taxonomy in \cite{prompt_eng_taxo}, used as a framework to select complementary prompting strategies that (i) define role and task boundaries, (ii) enforce strict grounding on repository artifacts, and (iii) standardize the output format for comparability.

Prompt development followed an iterative process. Successive variants were evaluated in parallel on ChatGPT, Gemini, and Qwen to identify recurring failure modes (e.g., incomplete answers, over-generalization, unsupported claims) and avoid model-specific overfitting. Refinement was conducted using repositories distinct from the final experimental dataset, progressively strengthening grounding constraints and stabilizing the response schema. Once consistent cross-model behavior was observed, the prompt was frozen to ensure fair comparison.

\begin{figure}[h!]
  \centering
  \includegraphics[width=\linewidth]{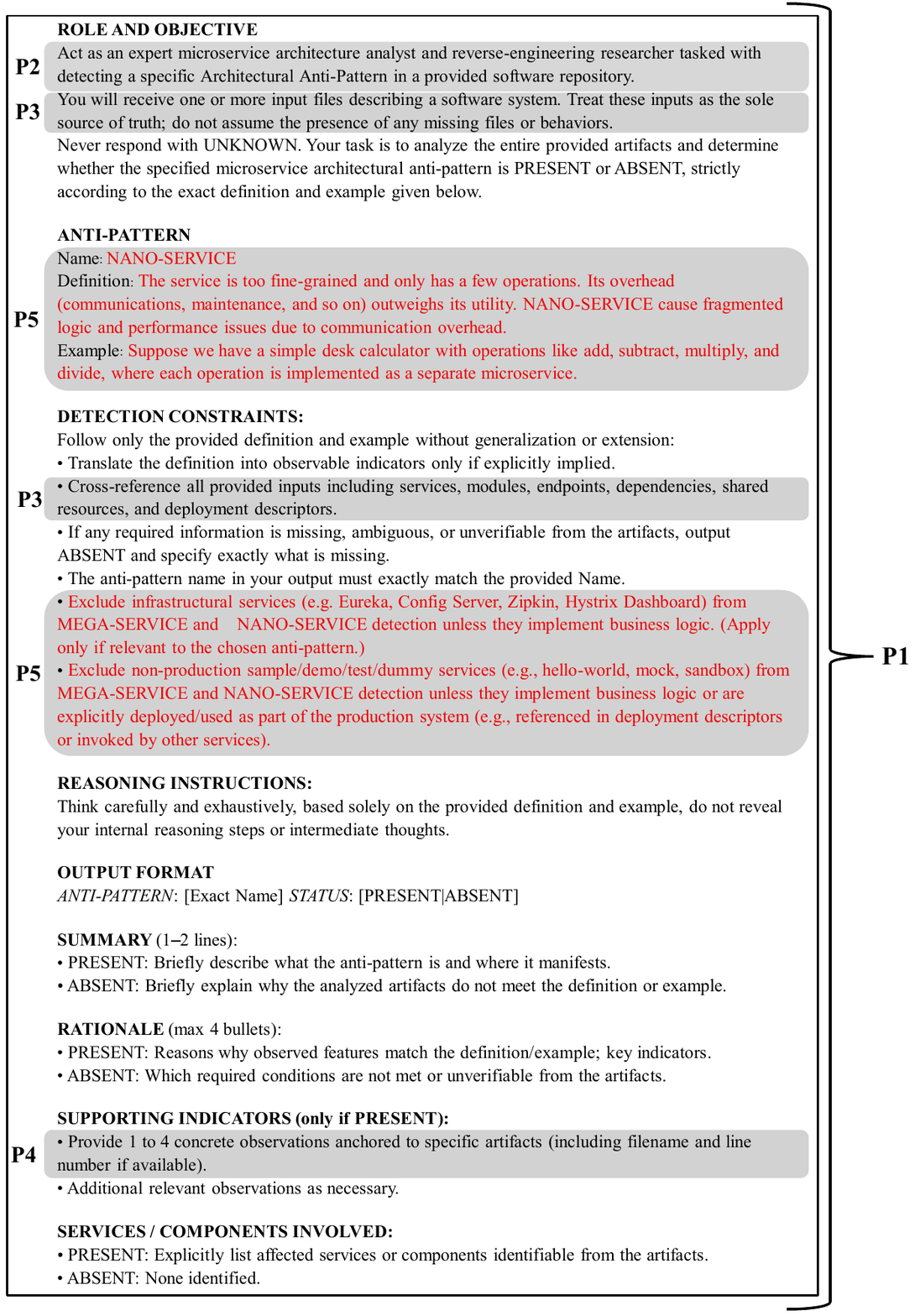}
  \caption{Instance of the parametric AP-detection prompt for \textit{Nano-service}, highlighting the shared template (black), the AP-dependent section (red), and the adopted prompting techniques (P$_x$).}
  \label{fig:workflow}
\end{figure}

The final prompt is organized into five sections, namely: \textit{Role and Objective, Anti-Pattern, Detection Constraints, Reasoning Instructions, and Output Format}.
Figure ~\ref{fig:workflow} shows an instantiated version of the prompt for detecting the Nano-service anti-pattern, highlighting the shared template (black) and the AP-specific section (red). In the prompt template, \textit{Role and Objective} assigns a specialized persona (microservice architecture and reverse-engineering expert) and defines task scope. \textit{Anti-Pattern} provides name, definition, and literature-based examples \cite{Cerny2023MicroserviceAntiPatterns, Tighilt2023}.
\textit{Detection Constraints} specifies observable evidence and enforces general and AP-specific rules. \textit{Reasoning Instructions} limits verbosity and discourages extended step-by-step reasoning to improve comparability. \textit{Output Format} prescribes a structured schema including decision (PRESENT/NOT PRESENT), involved services, and bounded supporting evidence.

The final design integrates five Prompt Engineering techniques:

\begin{enumerate}[label=$P\arabic*$:, labelsep=0.2em]

\item \textit{Gradual execution and output} \cite{prompt_eng_taxo}: the prompt is structured into explicit functional blocks to enhance consistency and automated result extraction.

\item \textit{Role-Prompting} \cite{role_prompt}: align terminology and focus on the analytical task.

\item \textit{Context management for improved interaction} \cite{prompt_eng_taxo}: the provided repository artifacts are treated as the sole source of truth, the model is forbidden to assume missing files or behaviors, and it is required to cross-reference relevant elements across the repository. This reduces over-generalization and discourages hallucinated evidence by keeping the analysis anchored to observable inputs %

\item \textit{Fact checklist} \cite{prompt_eng_taxo}: for \texttt{PRESENT} cases, the output must report a short list of concrete indicators grounded in repository artifacts (file and line number when available) and explicitly name the involved services/components, improving traceability and verifiability of detections. %

\item \textit{Example-based few-shot prompting} \cite{prompt_eng_taxo}: for each anti-pattern, the prompt provides its definition and one or more examples to calibrate the intended interpretation and decision boundary, improving consistency and reducing ambiguity across anti-pattern types.
\end{enumerate}

\section{Experimental Results}
\label{sec:results}

In this section we present the experimental results of our study on LLM-based anti-pattern detection in MSAs. For each RQ we report the observed results, provide an interpretation of them, and conclude reporting key findings and our answer to the RQ. 

When interpreting the results, special care is needed for low-frequency APs. Several anti-patterns have only a small number of GT instances; therefore, a single different prediction can substantially change the corresponding precision or recall value. For this reason, we report both ratios and percentages in Table~\ref{tab:ap-precision-recall}, and we base our conclusions on recurring trends across AP categories, models, and comparison scenarios rather than on isolated values for rare APs.

The supplemental materials, including the Ground Truth used in the study, the prompts adopted, and all outputs generated by the evaluated LLMs, are available online\footnote{\url{https://anonymous.4open.science/r/llm_ap_detecion-E5DF/README.md}}.

\subsection{RQ1 - Effectiveness of LLMs in Detecting MSA Anti-patterns}

RQ1 investigates how effectively different LLMs detect architectural anti-patterns in Microservice Architectures (MSAs). 
The first twelve columns of Table~\ref{tab:ap-precision-recall} report the precision and recall achieved by ChatGPT, Qwen, and Gemini across the 16 considered anti-patterns. 
For each anti-pattern, the highest precision value is \underline{underlined}, while the highest recall value is shown in \textbf{bold}.

\noindent \textbf{Results:}
Overall, the results reveal heterogeneous performance across both models and anti-pattern types. To better interpret these findings, we group anti-patterns into the three classes below.

A first class includes anti-patterns that were systematically undetected by all models. In particular, \textbf{CD} (Cyclic Dependency) falls into this group. None of the evaluated LLMs was able to correctly identify this anti-pattern. For instance, \textbf{CD} yields 0\% precision and 0\% recall for all three models, suggesting that it requires forms of architectural reasoning or dependency analysis that are difficult to infer from the information provided to the models.

A second subset of anti-patterns is detected by all models. This group includes \textbf{NAV} (No API Versioning), \textbf{NAG} (No API Gateway), and \textbf{NCI} (No CI/CD). For these anti-patterns, all LLMs achieve strong precision, indicating that the underlying architectural issues are easy to identify. For example, \textbf{NAV} reaches precision values between 75\% and 83\%, with recall between 90\% and 100\%; \textbf{NAG} and \textbf{NCI} are even more stable in precision, with all three models achieving 100\% precision.

The majority of anti-patterns fall into a third class, where detection effectiveness varies substantially across models. Examples include \textbf{WC} (Wrong Cuts), \textbf{NS} (Nano Service), and \textbf{MS} (Mega Service), for which different LLMs achieve markedly different levels of precision and recall. A representative case is \textbf{WC}: Qwen attains 100\% precision but only 20\% recall, whereas Gemini reaches 100\% recall at the cost of a much lower precision (42\%); ChatGPT instead shows a more balanced behavior with 100\% precision and 80\% recall. Similar variability is observed for other anti-patterns such as \textbf{SP} (Shared Persistence), \textbf{HE} (Hardcoded Endpoint), \textbf{TO}, \textbf{NHC}, \textbf{MSIPH}, \textbf{MC}, \textbf{IM}, and \textbf{LL}. For instance, in \textbf{HE}, ChatGPT achieves a precision of 74\% and a recall of 93\%, clearly outperforming Qwen, which reaches 56\% precision and 60\% recall, and Gemini, which achieves 63\% precision and 67\% recall. These results suggest that detecting such anti-patterns requires more nuanced architectural interpretation, particularly related to service granularity, system decomposition, and cross-service interactions. A representative example is \textbf{SP}, where models may correctly recover the known instances but also report additional candidates that are only apparently plausible. This typically happens when shared database names, similar connection settings, or repeated persistence-related artifacts suggest a data-sharing violation even when the services do not actually break data ownership boundaries. In the analyzed \textit{Apollo} repository, for instance, \texttt{apollo-configservice} and \texttt{apollo-adminservice} both access \texttt{ApolloConfigDB}. At first sight, this may look like a clear Shared Persistence case. However, the two services still appear to belong to the same broader configuration-management context, making the case more borderline than a canonical SP instance. This kind of situation helps explain how false positives can emerge: the repository contains architectural signals that are suggestive, but not sufficient on their own to confirm the anti-pattern.

Finally, considering the models individually, Qwen tends to achieve very high precision in several cases, although this often comes at the cost of lower recall. For example, it reaches 100\% precision on anti-patterns such as \textbf{NS}, \textbf{MS}, \textbf{WC}, and \textbf{TO}, but with recall dropping to 33\%, 50\%, 20\%, and 50\%, respectively. ChatGPT generally shows higher recall values, suggesting a greater tendency to identify potential anti-patterns, even at the risk of introducing false positives. Gemini exhibits more variable behavior, alternating between strong results, such as \textbf{NAV} with 100\% recall, and more selective profiles, such as \textbf{NHC}, where it reaches 100\% precision but only 33\% recall.

\noindent \textbf{Discussion:}
The observed results highlight several important characteristics of LLM-based architectural anti-pattern detection. Regarding the detectability of APs in MSAs, we observed the following trends.

First, certain anti-patterns appear difficult for LLMs to detect. In particular, the complete failure on CD and SL suggests that some architectural issues require structural analysis capabilities that are not easily inferred from the information provided to the models. 
Repository flattening may have contributed to this limitation by presenting dependency-related evidence as dispersed textual fragments rather than as explicit architectural relations. However, we do not consider flattening the only cause: these APs intrinsically require evidence about service boundaries, dependency directions, and cross-service relations that static repository artifacts may expose only partially.

Second, the consistent detection of NAV, NAG, and NCI suggests that some anti-patterns are more easily recognizable by LLMs. These issues are typically associated with missing architectural elements or development practices and may be easier to infer from service interfaces, configuration files, or project documentation.

Third, the mixed behavior observed across several anti-patterns suggests that LLM-based detection depends on how clearly the anti-pattern is reflected in the available artifacts. Anti-patterns that leave clear architectural signals tend to be detected more reliably, whereas those requiring deeper structural reasoning or cross-service analysis remain more challenging. This is especially true for issues related to service granularity or system decomposition, such as Wrong Cuts, Nano Service, and Mega Service. Detecting these cases requires judging architectural design choices rather than identifying missing components or explicit structural relations, which explains the greater variability across models.

Finally,  the evaluation also highlighted systematic behavioral differences among the considered LLMs. Qwen tends to produce highly accurate predictions, suggesting a more conservative detection strategy in which anti-patterns are reported only when the model is highly confident. In contrast, ChatGPT frequently achieves higher recall values, indicating a greater tendency to identify potential anti-patterns, even at the cost of introducing false positives. Gemini exhibits more variable behavior, with performance depending on the specific anti-pattern. These differences suggest that LLMs may adopt distinct strategies when reasoning about architectural descriptions, which can influence their suitability for different detection scenarios.

\noindent \textbf{Findings:} From the analysis, we derive the following findings:

\begin{enumerate}[label=\textbf{F\arabic*:}, leftmargin=*, itemsep=0.25em, labelsep=0.2em]
\item \textit{LLMs are effective when anti-patterns can be inferred from visible and semantically meaningful repository cues, especially when detection requires qualitative architectural judgment rather than rigid rule matching. This is evident for cases such as \textbf{WC}, \textbf{NS}, \textbf{NAV}, and \textbf{HE}, where models can leverage naming, organization, interface conventions, and scattered configuration hints to identify plausible architectural issues.}

\item \textit{LLMs remain weak on anti-patterns that require structural proof or exact technical verification, rather than plausible interpretation. This limitation emerges clearly for \textbf{CD}, \textbf{SL}, and \textbf{TO}, where reliable detection depends on demonstrating service-level cycles, distinguishing problematic library sharing from acceptable reuse, or extracting precise timeout-related configuration values.}

\item \textit{A recurrent LLM error is over-interpretation: models often treat suggestive architectural signals as sufficient evidence even when the surrounding context is ambiguous. This explains several false positives in anti-patterns such as \textbf{SP}, \textbf{MC}, and \textbf{MSIPH}, where shared resources, repeated configuration artifacts, or deployment clues may look suspicious without necessarily indicating a real anti-pattern.}

\item  \textit{LLM performance is strongly influenced by the type of missing or implicit evidence an anti-pattern requires. When the absence of an expected mechanism remains clearly visible in the repository, as in \textbf{NAG}, \textbf{NCI}, \textbf{IM}, and \textbf{NHC}, models can be competitive; however, performance becomes less reliable when absence detection requires deeper contextual or operational inference, as in \textbf{LL}.}
\end{enumerate}

The observations discussed above can be summarized into the following direct answer to RQ1:

\begin{tcolorbox}[colback=blue!10,boxrule=0.5pt,title=RQ1 Answer,boxsep=1pt,left=1pt,right=1pt,top=1pt,bottom=1pt,before skip=2pt, after skip=2pt]
LLMs show heterogeneous effectiveness in detecting MSA architectural anti-patterns. They are more effective when anti-patterns can be inferred from visible and semantically meaningful repository cues, but they remain weak when detection requires structural proof or deeper cross-service reasoning. Some anti-patterns are consistently missed by all models, whereas others are consistently detected across different LLMs. For many cases, however, effectiveness remains strongly model-dependent, confirming that current LLMs can support anti-pattern detection but are not yet robust across all anti-pattern types.
\end{tcolorbox}

\begin{table*}[h!]

\caption{Precision and recall per AP for \textbf{ChatGPT}, \textbf{Qwen}, \textbf{Gemini}, and \textbf{MARS}.  \\ \textbf{Legend:} AP: Anti-Pattern. \textbf{Character Code:} Best precision values (\%) per row are underlined; best recall values (\%) per row are in bold.}
\centering
\label{tab:ap-precision-recall}
\small
\renewcommand{\arraystretch}{0.9}
\setlength{\tabcolsep}{4pt}
\color{black}
\resizebox{0.9\linewidth}{!}{%
\begin{tabular}{c c c c c c c c c c c c c c c c c}
\toprule
\multirow{3}{*}{\textbf{AP}} &
\multicolumn{4}{c}{\textbf{ChatGPT}} &
\multicolumn{4}{c}{\textbf{Qwen}} &
\multicolumn{4}{c}{\textbf{Gemini}} &
\multicolumn{4}{c}{\textbf{MARS}} \\
\cmidrule(lr){2-5}\cmidrule(lr){6-9}\cmidrule(lr){10-13}\cmidrule(lr){14-17}
& \multicolumn{2}{c}{\textbf{Precision}} & \multicolumn{2}{c}{\textbf{Recall}}
& \multicolumn{2}{c}{\textbf{Precision}} & \multicolumn{2}{c}{\textbf{Recall}}
& \multicolumn{2}{c}{\textbf{Precision}} & \multicolumn{2}{c}{\textbf{Recall}}
& \multicolumn{2}{c}{\textbf{Precision}} & \multicolumn{2}{c}{\textbf{Recall}} \\
\cmidrule(lr){2-3}\cmidrule(lr){4-5}\cmidrule(lr){6-7}\cmidrule(lr){8-9}\cmidrule(lr){10-11}\cmidrule(lr){12-13}\cmidrule(lr){14-15}\cmidrule(lr){16-17}
& \textbf{Ratio} & \textbf{\%} & \textbf{Ratio} & \textbf{\%}
& \textbf{Ratio} & \textbf{\%} & \textbf{Ratio} & \textbf{\%}
& \textbf{Ratio} & \textbf{\%} & \textbf{Ratio} & \textbf{\%}
& \textbf{Ratio} & \textbf{\%} & \textbf{Ratio} & \textbf{\%} \\
\midrule

\multicolumn{17}{c}{\textbf{Intra-Service Design}} \\
\midrule
NS    & 1/3   & 33\% & 1/3   & 33\% & 1/1   & \underline{100\%} & 1/3   & 33\% & 2/7   & 29\% & 2/3   & \textcolor{black}{\textbf{67\%}} & 0/5   & 0\%  & 0/3   & 0\% \\
MS    & 1/2   & 50\% & 1/4   & 25\% & 2/2   & \underline{100\%} & 2/4   & 50\% & 2/3   & 67\% & 2/4   & 50\% & 4/6   & 67\% & 4/4   & \textcolor{black}{\textbf{100\%}} \\
NAV   & 10/13 & 77\% & 10/10 & \textcolor{black}{\textbf{100\%}} & 9/12  & 75\%  & 9/10  & 90\% & 10/12 & 83\% & 10/10 & \textcolor{black}{\textbf{100\%}} & 7/8   & \underline{88\%} & 7/10  & 70\% \\

\midrule
\multicolumn{17}{c}{\textbf{Inter-Service Decomposition}} \\
\midrule
CD    & 0/1   & \underline{0\%}  & 0/5   & \textcolor{black}{\textbf{0\%}}  & 0/1   & \underline{0\%}   & 0/5   & \textcolor{black}{\textbf{0\%}}  & 0/0   & \underline{0\%}  & 0/5   & \textcolor{black}{\textbf{0\%}}  & 0/0     & \underline{0\%}    & 0/5     & \textbf{0\%}    \\
SP    & 1/4   & 25\% & 1/1   & \textcolor{black}{\textbf{100\%}} & 1/3   & 33\%  & 1/1   & \textcolor{black}{\textbf{100\%}} & 1/2   & \underline{50\%} & 1/1   & \textcolor{black}{\textbf{100\%}} & 1/2   & \underline{50\%} & 1/1   & \textcolor{black}{\textbf{100\%}} \\
SL    & 0/1   & 0\%  & 0/2   & 0\%  & 0/1   & 0\%   & 0/2   & 0\%  & 0/3   & 0\%  & 0/2   & 0\%  & 2/2   & \underline{100\%} & 2/2   & \textcolor{black}{\textbf{100\%}} \\
WC    & 4/4   & \underline{100\%} & 4/5   & 80\% & 1/1   & \underline{100\%} & 1/5   & 20\% & 5/12  & 42\% & 5/5   & \textcolor{black}{\textbf{100\%}} & 3/3   & \underline{100\%} & 3/5   & 60\% \\

\midrule
\multicolumn{17}{c}{\textbf{Service Interaction}} \\
\midrule
HE    & 14/19 & \underline{74\%} & 14/15 & \textcolor{black}{\textbf{93\%}} & 9/16  & 56\%  & 9/15  & 60\% & 10/16 & 63\% & 10/15 & 67\% & 6/10  & 60\% & 6/15  & 40\% \\
NAG   & 3/3   & \underline{100\%} & 3/5   & 60\% & 4/4   & \underline{100\%} & 4/5   & 80\% & 4/4   & \underline{100\%} & 4/5   & 80\% & 5/9   & 56\% & 5/5   & \textcolor{black}{\textbf{100\%}} \\
TO    & 2/4   & 50\% & 2/4   & 50\% & 2/2   & \underline{100\%} & 2/4   & 50\% & 2/7   & 29\% & 2/4   & 50\% & 3/3   & \underline{100\%} & 3/4   & \textcolor{black}{\textbf{75\%}} \\
NHC   & 4/6   & 67\% & 4/6   & \textcolor{black}{\textbf{67\%}} & 2/4   & 50\%  & 2/6   & 33\% & 2/2   & \underline{100\%} & 2/6   & 33\% & 3/5   & 60\% & 3/6   & 50\% \\

\midrule
\multicolumn{17}{c}{\textbf{Team Organization}} \\
\midrule
MSIPH & 3/9   & 33\% & 3/4   & \textcolor{black}{\textbf{75\%}} & 3/6   & 50\%  & 3/4   & \textcolor{black}{\textbf{75\%}} & 3/9   & 33\% & 3/4   & \textcolor{black}{\textbf{75\%}} & 3/3   & \underline{100\%} & 3/4   & \textcolor{black}{\textbf{75\%}} \\
NCI   & 8/8   & \underline{100\%} & 8/11  & \textcolor{black}{\textbf{73\%}} & 7/7   & \underline{100\%} & 7/11  & 64\% & 8/8   & \underline{100\%} & 8/11  & \textcolor{black}{\textbf{73\%}} & 8/8   & \underline{100\%} & 8/11  & \textcolor{black}{\textbf{73\%}} \\
MC    & 5/8   & 63\% & 5/5   & \textcolor{black}{\textbf{100\%}} & 3/3   & \underline{100\%} & 3/5   & 60\% & 2/2   & \underline{100\%} & 2/5   & 40\% & 5/8   & 63\% & 5/5   & \textcolor{black}{\textbf{100\%}} \\
IM    & 1/1   & \underline{100\%} & 1/3   & 33\% & 1/1   & \underline{100\%} & 1/3   & 33\% & 1/1   & \underline{100\%} & 1/3   & 33\% & 3/8   & 38\% & 3/3   & \textcolor{black}{\textbf{100\%}} \\
LL    & 1/4   & 25\% & 1/4   & 25\% & 2/8   & 25\%  & 2/4   & 50\% & 2/8   & 25\% & 2/4   & 50\% & 4/6   & \underline{67\%} & 4/4   & \textcolor{black}{\textbf{100\%}} \\

\bottomrule
\end{tabular}%
}
\end{table*}

\subsection{RQ2 - Comparison Between LLMs and MARS}

RQ2 investigates how LLM-based approaches compare against a state-of-the-art static analysis tool (MARS) for detecting architectural anti-patterns in MSAs. The last four columns of Table \ref{tab:ap-precision-recall} report precision and recall values for MARS across the considered anti-patterns.

\noindent \textbf{Results:}
The results show that detection performance differs substantially depending on the anti-pattern considered. Overall, three main scenarios emerge.

First, there is a set of anti-patterns where \textit{MARS clearly outperforms all LLMs}. This includes cases where MARS detects all occurrences while LLMs miss a significant portion of them, and even situations where LLMs fail to detect the anti-pattern entirely. Examples of this behavior include \textbf{SL}, where all LLMs achieve 0\% precision and 0\% recall while MARS reaches 100\% precision and 100\% recall, and \textbf{MS}, where MARS achieves 100\% recall compared with a best LLM recall of only 50\%. A similar gap also appears for \textbf{LL}, where MARS reaches 67\% precision and 100\% recall, whereas the best LLM recall stops at 50\%, and for \textbf{IM}, where all LLMs achieve only 33\% recall while MARS reaches 100\%. A similar trend is also observed for \textbf{NAG} and \textbf{TO}, where MARS achieves 100\% and 75\% recall, respectively, compared with a best LLM recall of 80\% and 50\%. In addition, for \textbf{MSIPH} the recall values are comparable across approaches (75\%), but MARS shows higher precision (100\% vs. at most 50\%).

Second, there are anti-patterns where \textit{LLMs outperform MARS}. In particular, LLMs show stronger detection capabilities for \textbf{NAV}, \textbf{HE}, and \textbf{NS}. For \textbf{NAV}, both ChatGPT and Gemini achieve 100\% recall, outperforming MARS at 70\%. For \textbf{HE}, ChatGPT reaches 74\% precision and 93\% recall, clearly improving over MARS, which stops at 60\% precision and 40\% recall. For \textbf{NS}, Gemini achieves 67\% recall, while MARS does not detect any occurrence (0\% recall).

Third, a set of anti-patterns shows \textit{comparable performance between the two approaches}. For example, \textbf{NCI} exhibits identical precision across all approaches (100\%) and very similar recall values, with ChatGPT, Gemini, and MARS all reaching 73\%. Comparable behavior is also observed for \textbf{MC}, where ChatGPT and MARS both achieve 63\% precision and 100\% recall, and for \textbf{SP}, where Gemini and MARS both reach 50\% precision and 100\% recall. \textbf{NHC} is also relatively close, with ChatGPT achieving 67\% precision and 67\% recall, compared with 60\% precision and 50\% recall for MARS. For \textbf{CD}, none of the LLMs detect occurrences of the anti-pattern, and MARS does not provide a useful advantage either. Finally, \textbf{WC} shows heterogeneous behavior, with ChatGPT and MARS both reaching 100\% precision, ChatGPT improving recall to 80\%, and Gemini achieving the highest recall overall at 100\% but with much lower precision (42\%).

\noindent \textbf{Discussion:}
The results highlight complementary strengths between LLM-based detection and static analysis. MARS performs better when anti-patterns can be identified through explicit structural properties, such as dependency violations, architectural constraints, or service relationships, which likely explains its stronger performance on \textbf{SL}, \textbf{MS}, \textbf{LL}, and \textbf{IM}. By contrast, LLMs perform better when detection requires semantic interpretation of the system design or developer intent, which may explain their stronger results on \textbf{NAV} and \textbf{HE}. For several anti-patterns, however, both approaches achieve comparable results, suggesting that some architectural issues can be detected both through structural analysis and through semantic reasoning over repository artifacts.

\noindent \textbf{Findings:}
Overall, these findings indicate that LLMs and static analysis tools rely on different detection mechanisms and therefore exhibit complementary capabilities. Rather than replacing traditional analysis tools, LLMs may provide additional support by identifying anti-patterns that are harder to capture with purely structural rules.

\begin{enumerate}[label=\textbf{F\arabic*:}, leftmargin=*, itemsep=0.2em, labelsep=0.2em]
\item \textit{LLMs tend to outperform or complement static analysis when anti-pattern detection requires semantic interpretation and qualitative architectural judgment rather than rule matching. This is evident in cases such as \textbf{WC}, \textbf{NS}, \textbf{NAV}, and especially \textbf{HE}, where models can combine heterogeneous repository cues, naming conventions, interface structure, and scattered configuration hints in ways that are hard to encode through fixed static rules alone.}
    
\item  \textit{State-of-the-art static analysis tools retain a clear advantage when detection depends on structural proof, exact technical verification, or systematically extractable indicators. This limit of LLMs emerges in cases such as \textbf{CD}, \textbf{SL}, \textbf{MS}, and \textbf{TO}, where detection requires proving service-level cycles, distinguishing problematic dependency sharing from acceptable reuse, approximating service size through explicit thresholds, or identifying precise timeout-related parameters.}
    
\item \textit{The main trade-off between LLMs and static analysis is interpretive flexibility versus detection discipline. LLMs can infer plausible anti-patterns from partial, distributed evidence, but this flexibility often leads to over-interpretation and false positives, as in \textbf{SP}, \textbf{MC}, and \textbf{MSIPH}. Static analysis is more constrained and often more selective, especially for absence-based or configuration-driven anti-patterns such as \textbf{NAG}, \textbf{NCI}, \textbf{NHC}, \textbf{IM}, and \textbf{LL}.}
\end{enumerate}

The above findings can be summarized into the following overall answer to RQ2: 

\begin{tcolorbox}[colback=blue!10,boxrule=0.5pt,title=RQ2 Answer,boxsep=1pt,left=1pt,right=1pt,top=1pt,bottom=1pt,before skip=2pt, after skip=2pt] 
LLMs and the static analysis tool MARS exhibit complementary strengths in detecting MSA architectural anti-patterns. 
MARS performs better when detection depends on explicit structural evidence and verifiable architectural properties, whereas LLMs show advantages when detection requires semantic interpretation and qualitative judgment. For several anti-patterns, the two approaches achieve comparable performance. 
Overall, LLMs are neither consistently better nor worse than state-of-the-art static analysis tools: their effectiveness depends on the target anti-pattern and the evidence needed to detect it.
\end{tcolorbox}

\section{AP Category Takeaways}
\label{sec:ap-category-takeaways}

This section discusses the results of RQ1 and RQ2 through the AP categories considered in our study (Table~\ref{tab:ap_list}).
Rather than analyzing anti-patterns in isolation, we synthesize the main insights emerging within each category to highlight where LLM-based detection is more effective, where it remains limited, and how these trends compare with the static baseline \textbf{MARS}. This category-based discussion makes the practical implications of our findings more explicit.

\subsection{Inter-service decomposition}


Inter-service decomposition is the category in which prompt-based inspection over flattened repositories is most clearly limited. Flattening preserves source code, configuration files, and repository structure, but it linearizes the evidence and does not provide explicit architectural representations of service boundaries, dependency directions, deployment relations, or runtime interactions. As a result, it may exacerbate LLM difficulties for APs that require reasoning across service boundaries, although the underlying challenge is also due to the limited availability of explicit structural evidence in static artifacts. This is why \textbf{WC}, \textbf{CD}, \textbf{SL}, and \textbf{SP} show the widest gap between what LLMs can infer from repository-level cues and what MARS can verify through explicit checks. In \textbf{WC}, LLMs can detect broad signs of technically layered decomposition, but may over-interpret repository organization and misclassify legitimate supporting services as wrong decomposition. \textbf{CD} is an even clearer failure case, since detecting cyclic dependencies requires proving service-level cycles rather than spotting suggestive local clues. \textbf{SL} follows the same logic: distinguishing problematic internal library sharing from acceptable dependency reuse requires more explicit dependency-level evidence than flattened inputs usually expose. \textbf{SP} is slightly different, because all approaches recover the known positives, but LLMs tend to be less selective and more easily triggered by repeated database-related cues when bounded-context separation is unclear.

\noindent\textbf{Category Takeaway.} \textit{Inter-service decomposition is the category in which repository flattening is least sufficient: LLMs can often recognize broad decomposition symptoms, but they remain weak when detection requires explicit proof of cross-service structure or dependency relations.}

\subsection{Intra-service design}

In the \textit{Intra-service design} category, LLMs appear most effective. Compared with inter-service decomposition, these APs depend more on service-local evidence and on interpreting a service’s role, scope, and internal organization. This is especially visible in \textbf{NS}, where MARS retrieves no positives while the LLMs recover part of the GT, suggesting that identifying overly fine-grained services benefits from qualitative judgment over cues such as service thinness, sparse internal structure, and narrowly scoped responsibilities. \textbf{NAV} is another favorable case for LLMs: versioning evidence is often distributed across routes and configuration artifacts, and the models seem able to aggregate these heterogeneous local cues effectively. By contrast, \textbf{MS} remains more favorable to MARS, since oversized services can be approximated more robustly through explicit size- and structure-related indicators, whereas LLMs must infer whether implementation breadth truly reflects excessive business scope. Overall, this category shows that LLMs are strongest when the AP can be inferred from semantically meaningful local evidence rather than from rigid structural thresholds alone.

\noindent\textbf{Category Takeaway.} \textit{Intra-service design is the category in which LLMs are most competitive, especially when AP detection depends on local evidence and qualitative architectural judgment rather than on rigid structural proxies.}

\subsection{Service interaction}
The service-interaction category shows a more mixed picture. Here, performance depends strongly on the evidence required by each AP. Some anti-patterns remain accessible to LLM-based reasoning because they can be inferred from semantically meaningful and fairly local clues. This is the case of \textbf{HE}, where ChatGPT in particular appears able to go beyond narrow syntactic forms and identify endpoint hard-coding from scattered evidence across constants and configuration access. \textbf{NAG} also shows that LLMs can be competitive, although they tend to be more cautious than MARS when the task requires inferring the absence of gateway support. At the same time, the category includes cases in which MARS retains a clear advantage. \textbf{TO} depends on extracting exact timeout-related parameters from configuration artifacts, which is better suited to static analysis than to prompt-based interpretation. \textbf{NHC} is more model-dependent and highlights a broader absence-detection issue: some models remain competitive, but performance varies with how confidently the model infers that an expected mechanism is missing. Overall, this category shows that service-interaction APs are split between interpretable local cases and configuration-intensive or absence-based cases, with the latter remaining harder for LLMs.

\noindent\textbf{Category Takeaway.} \textit{Service-interaction APs split into two groups: LLMs remain competitive when evidence is semantically interpretable and relatively local, but they lose ground when detection depends on exact configuration extraction or on reliably proving the absence of expected mechanisms.}

\subsection{Team organization}

These APs are closely tied to DevOps artifacts, deployment conventions, and operational semantics, and therefore show a recurring trade-off between visibility of evidence and interpretability. Some remain favorable to both LLMs and static analysis because the decisive evidence is explicit and localized in standard artifacts. \textbf{NCI} is the clearest case, since CI/CD adoption is usually reflected in highly recognizable files, making the AP visible even after repository flattening. \textbf{MC} is also detectable by all approaches, although broader repository exploration tends to improve recall at the cost of more false positives. Other APs are harder for LLMs because recognizing local artifacts is not enough to determine whether the architectural condition truly holds. \textbf{LL} is the clearest example: identifying logging-related code or configuration does not suffice to determine whether logging is properly centralized. \textbf{MSIPH} follows a similar logic, as repository-only inputs often provide partial deployment clues but not enough evidence to distinguish problematic co-location from acceptable operational choices. Finally, \textbf{IM} shows a typical absence-detection trade-off: LLMs are selective and precise, whereas MARS improves recall through systematic checks but with more false positives.

\noindent\textbf{Category Takeaway.} \textit{For this category LLMs perform well when the evidence is explicit and localized in standard repository artifacts, but they degrade when the judgment depends on deployment semantics, infrastructure interpretation, or proving that an operational capability is missing.}

\subsection{Final discussion}

We now synthesize the category-specific observations into a set of broader \textit{Final Takeaways} (\textbf{FT}) on the overall behavior of the evaluated LLMs and of \textbf{MARS}.

\noindent\textbf{FT1.} \textit{The main factor that explains the relative advantage of LLMs or MARS is the type of evidence required by the anti-pattern.} LLMs are strongest when the decisive evidence is local, semantically rich, and directly visible in the repository, as in \textbf{NS}, \textbf{NAV}, \textbf{NCI}, and partly \textbf{HE}. By contrast, \textbf{MARS} is stronger when detection depends on explicit cross-service structure, precise configuration parameters, or deployment-level wiring, as in \textbf{CD}, \textbf{SL}, \textbf{TO}, \textbf{LL}, and partly \textbf{MSIPH}.

\noindent\textbf{FT2.} \textit{The three LLMs exhibit recurring behavioral profiles rather than random differences.} \textbf{Qwen} is generally more conservative and often gains precision at the cost of recall. \textbf{Gemini} is more easily triggered by broad or ambiguous architectural symptoms and is therefore more exposed to false positives in borderline cases, as shown by \textbf{WC}. \textbf{ChatGPT} most often occupies a middle ground, although in some APs, such as \textbf{SP}, it can still apply the AP definition too mechanically.

\noindent\textbf{FT3.} \textit{A central limitation of flattened-repository prompting is not only missing evidence, but fragmented evidence.} The models can often recognize local indicators, but they are less robust when the judgment requires connecting clues distributed across services, configuration files, dependency descriptors, and infrastructure artifacts. \textbf{CD} and \textbf{SL} are the clearest examples, but the same issue also emerges in \textbf{SP}, \textbf{LL}, and \textbf{MSIPH}.

\noindent\textbf{FT4.} \textit{The main lesson learned is hybrid rather than competitive.} Static analysis should remain the backbone when an AP can be identified through explicit, verifiable facts. LLM-based inspection is most useful as a complement for APs that require architectural interpretation or rely on heterogeneous, semantically meaningful cues that are hard to encode as rigid rules. The most promising direction is therefore not prompt-only replacement, but hybrid pipelines combining LLMs with richer structural inputs such as dependency graphs, build relations, and service interaction data. 

\section{Threats to Validity}
\label{sec:threats}

We discuss the main threats to validity that may affect our findings.

\noindent \textbf{Internal Validity.}
This category concerns factors that may have influenced the observed results.
First, LLM performance may be affected by prompt design. Although we adopted a structured, parametric prompt and froze it after iterative refinement on repositories distinct from the evaluation dataset, different prompt formulations might lead to different detection behavior. Second, repository flattening may have altered structural signals. While \textsc{Repomix} was configured to retain source code and configuration files only, transforming repositories into a single textual artifact may have reduced the explicit visibility of cross-service relationships, potentially affecting LLM reasoning and the comparison with MARS. Third, LLM outputs are inherently stochastic. Although we followed a uniform execution protocol across models, minor generation differences could still affect borderline cases.

Moreover, as a limited robustness check, we repeated selected low-frequency repository-AP inputs five times under the same prompt and input conditions, observing unchanged detection outcomes; however, since this check was not systematic across all repositories and APs, we do not use it as a full variance analysis

\noindent \textbf{Construct Validity.}
This threat concerns whether the adopted measures and evaluation choices accurately capture APs detection effectiveness. We rely on the GT provided in the original \textsc{MARS} study. Although this dataset is curated, the GT is not an independently created benchmark specifically designed for comparing LLMs and static-analysis tools. Therefore, benchmark bias may exist: the selected repositories and AP labels reflect the scope and operationalization of the original \textsc{MARS} study, and the restriction to the 16 \textsc{MARS}-detectable APs may underrepresent anti-patterns for which LLMs could provide stronger advantages. For this reason, we interpret our results as evidence about the behavior of LLMs on a MARS-derived benchmark, not as a definitive head-to-head evaluation of LLMs against static analysis in general.
In particular, some APs (e.g., NS or MS) involve qualitative judgments that may not be uniquely determined from repository artifacts alone. Moreover, the evaluation combines system-level and service-level detection under precision and recall. While these metrics capture detection performance, they do not reflect explanation quality or the practical usefulness of LLM outputs.

\noindent \textbf{Conclusion Validity.}
Our evaluation covers a limited number of repositories and 16 anti-patterns, some of which occur rarely in the benchmark. Consequently, precision and recall may vary substantially with small changes in predictions for low-frequency APs. We therefore interpret the results as descriptive evidence and focus on recurring trends rather than isolated values. We did not perform statistical hypothesis testing because the dataset size and structure do not support robust inferential analysis. Repository-level bootstrap resampling is limited by the small number of repositories, while AP-level resampling would violate independence assumptions. Likewise, McNemar's test is not well suited to our mixed repository- and service-level evaluation setting. Therefore, our conclusions should be interpreted as empirical trends within the studied benchmark rather than statistically validated evidence of superiority.

\noindent \textbf{External Validity.}
This threat concerns the generalizability of our results. Our findings are based on open-source microservice systems previously analyzed by MARS and may not generalize to industrial systems with different architectures, technology stacks, or organizational settings. We also evaluated only three general-purpose LLMs using static repository inputs, so results may change with other models, fine-tuned variants, or richer inputs such as runtime traces or dependency graphs. Further studies on larger and more diverse datasets are therefore needed.

\section{Conclusion and Future Work}
\label{sec:conclusion}

This paper investigated the use of LLMs for architectural anti-pattern detection in microservice architectures from static repository artifacts. By evaluating three general-purpose LLMs on a curated benchmark of microservice systems and comparing them with the state-of-the-art static-analysis tool MARS, we showed that LLMs can already provide useful support for this task, although their effectiveness varies substantially across anti-patterns. More specifically, our results indicate that the main factor shaping detection performance is the type of evidence required by the target anti-pattern. LLMs proved more effective when the relevant evidence is local, heterogeneous, and semantically interpretable, whereas they remained weaker on cases that require explicit structural, operational, or cross-service dependency information. Overall, our findings suggest that LLMs are not yet a replacement for traditional static analyzers, but rather a promising complementary aid for architectural assessment in microservice systems.

As future work, we plan to extend this study in several directions. First, we aim to enrich the input given to LLMs with more explicit architectural evidence, such as dependency graphs and service interaction data, to support the detection of APs that are currently difficult to identify from flattened repositories alone. Second, we intend to investigate hybrid approaches that combine static-analysis outputs with LLM-based reasoning. Third, we plan to broaden the empirical evaluation by including additional repositories, models, and industrial systems to strengthen the generalizability of the findings.

\section*{Acknowledgment}
This work was supported by the Italian Ministry of Research under the complementary actions to the NRRP “Fit4MedRob -- Fit for Medical Robotics” Grant (\#PNC0000007) and by the GATT project (GAmification in Testing Teaching), funded by the University of Naples Federico II Research Funding Program (FRA). The authors thank Master’s student Antimo Barbato for his contribution to the definition of the pipeline and the execution of the experiments.

\clearpage

\balance
\bibliographystyle{IEEEtran}
\bibliography{biblio}

\end{document}